\documentclass[12pt,a4paper]{article}

\usepackage[intlimits,tbtags]{amsmath}
\usepackage{amsfonts}
\usepackage{amssymb}
\usepackage{cite}
\usepackage{latexsym}
\usepackage{a4}

\newcommand{\I}{\mathrm{i}}

\newcommand{\tr}{\triangleright}

\newcommand{\R}{\mathcal{R}}

\newcommand{\op}{\mathrm{op}}
\newcommand{\adL}{{\mathrm{ad_L}}}

\newcommand{\Proj}{\mathbb{P}}
\newcommand{\Acal}{\mathcal{A}} 
\newcommand{\Hcal}{\mathcal{H}}
\newcommand{\Xcal}{\mathcal{X}}
\newcommand{\Scal}{\mathcal{S}}

\newcommand{\lrAngle}[1]{\langle #1\rangle}

\newcommand{\slq}{{\mathcal{U}_q(\mathrm{sl}_2)}}

\newcommand{\suq}{{\mathcal{U}_q(\mathrm{su}_2)}}
\newcommand{\slC}{{\mathcal{U}_q(\mathrm{sl}_2(\mathbb{C})) }}

\newcommand{\SUq}{{SU_q(2)}}

\newcommand{\Mink}{{\mathbb{R}_q^{1,3}}}

\begin{document}

\begin{center}

{\Large{\bf Free \textit{q}-Deformed Relativistic Wave Equations by
  Representation Theory}}

\vspace{3em}

\textbf{Christian Blohmann}

\vspace{1em}

Ludwig-Maximilians-Universit\"at M\"unchen, Sektion Physik\\
Lehrstuhl Prof.\ Wess, Theresienstr.\ 37, D-80333 M\"unchen\\[1em]

Max-Planck-Institut f\"ur Physik,
        F\"ohringer Ring 6, D-80805 M\"unchen\\[1em]

\end{center}

\vspace{1em}

\begin{abstract}
  In a representation theoretic approach a free $q$-relativistic wave
  equation must be such, that the space of solutions is an irreducible
  representation of the $q$-Poincar{\'e} algebra. It is shown how this
  requirement uniquely determines the $q$-wave equations. As examples,
  the $q$-Dirac equation (including $q$-gamma matrices which satisfy a
  $q$-Clifford algebra), the $q$-Weyl equations, and the $q$-Maxwell
  equations are computed explicitly.
\end{abstract}

\section{Introduction}

Quantum field theories on noncommutative spaces have been receiving a
tremendous amount of attention during the last few years (for reviews
see \cite{Douglas:2001,Szabo:2001}) and, indeed, considerable progress
was made.  Noncommutative geometry has naturally appeared in a certain
low energy limit of string theory \cite{Seiberg:1999} and gauge
theories on general noncommutative geometries have found a solid,
perturbative formulation \cite{Madore:2000b,Jurco:2001} within the
framework of deformation quantization.  Expanding the product of
noncommutative quantum fields perturbatively and relating the
noncommutative gauge potentials and fields to their ordinary,
commutative counterparts via the Seiberg-Witten map has put quantum
theories on noncommutative spaces within the range of phenomenological
considerations: A minimal noncommutative extension of the standard
model was formulated \cite{Calmet:2001}, the effects of noncommutative
geometry on magnetic and electric moments was studied
\cite{Iltan:2003,Minkowski:2003}, noncommutative neutrino-photon
coupling with possible astrophysical implications was investigated
\cite{Schupp:2002}, the OPAL collaboration has started looking for
noncommutative signatures in electron positron pair annihilation
\cite{Abbiendi:2003}, just to name some recent examples.  For a review
on the phenomenological implications of noncommutative geometry see
\cite{Hinchliffe:2002}.

Currently, most papers studied the particularly simple case where the
commutator of the space-time observables $[X_\mu, X_\nu] =
\theta_{\mu\nu}$ is a constant antisymmetric matrix. This kind of
noncommutativity can be viewed as due to a constant background field.
In string theory it can be attributed to a constant $B$-field on a
D-brane. Clearly, a noncommutativity, which originates from a constant
background field breaks Lorentz symmetry. It could be argued that if
the noncommutativity parameters $\theta_{\mu\nu}$ are small, the
violation of Lorentz symmetry is only small, too. However, on the
level of regularization of loop diagrams, the noncommutativity leads
to an interdependence of ultra-violet and infra-red cutoff scales
\cite{Minwalla:1999,Matusis:2000}.  Phenomenologically, this UV/IR
mixing is problematic, as it seems to put even large scale Lorentz
symmetry and weakened notions of locality of noncommutative quantum
field theory into doubt \cite{Alvarez-Gaume:2003}. But as yet, UV/IR
mixing was investigated in detail only for the case of constant
$\theta_{\mu\nu}$.

In a self-contained theory it would be reasonable to expect
$\theta_{\mu\nu}$ to become a dynamical quantity itself, which
transforms covariantly with respect to some (perhaps generalized)
space-time symmetry. Since constant $\theta_{\mu\nu}$ does not allow
for a perturbative deformation of Lorentz symmetry, one has to look
for alternatives. In this context, it has been proposed repeatedly
\cite{Madore:2000b,Jurco:2001} to investigate quantum spaces as
standard examples for noncommutative geometries with generalized
symmetries, quantum groups, which are controlled by the same parameters
as the noncommutativity of the spaces.  On quantum spaces, the changes
induced by noncommutativity to physical concepts that are tied to
space-time symmetry, such as energy-momentum conservation, Lorentz
invariance, independence of in and out states etc. would be better
integrated in the perturbative approach to noncommutative gauge
theories. This would be an advantage for phenomenological
considerations.

Another motivation to study noncommutative space-times with quantum
group symmetries has emerged from the attempts to explain the
observation \cite{Bird:1995} of cosmic rays of energy beyond the
spectral cutoff (the Greisen-Zatsepin-Kuzmin limit) which is expected
due to interaction with the cosmic microwave background. The often
proposed explanation of such ultra high energy rays by vacuum
dispersion relations, that is, the dependence of the speed of light on
the wavelength, was shown by Amelino-Camelia to be reconcilable in
principle with the observer independence of the laws of physics
\cite{Amelino-Camelia:2000}. This leads to a modification of special
relativity by the assumption that there is not only an observer
invariant velocity but also an observer invariant length, the Planck
length. This proposition, now called doubly special relativity, has
initiated a large number of active studies from both, the mathematical
and the phenomenological viewpoint. (For an overview see
\cite{Amelino-Camelia:2002}.) Remarkably, many of the concrete
realizations of doubly special relativity have led to quantum spaces
with a quantum group Lorentz symmetry \cite{Agostini:2003}.

Mathematically, the construction of noncommutative gauge theories on
quantum spaces is quite involved, since covariance with respect to the
quantum symmetry must be preserved at every step of the construction.
Previous work in dealing with $q$-deformations has studied
realizations of quantum spaces by star products
\cite{Wachter:2001,Blohmann:2002a} and the calculation of the
Seiberg-Witten map for Abelian gauge theory on the quantum plane in
first order \cite{Mesref:2002}. Ultimately, we would like to establish
the Feynman rules for QED on quantum Minkowski space. Towards this
goal, the free fermionic and bosonic propagators have to be
determined. Propagators can be viewed as inverses or Green's functions
of free wave equations. Hence, one of the first steps is to establish
the free wave equations for fermions and gauge bosons. This is the
purpose of the present paper.

Free elementary particles can be identified with irreducible
representations of the Poincar{\'e} group \cite{Wigner:1939} or,
equivalently, the Poincar{\'e} algebra. The representations are
realized by Wigner spinors, that is, on-shell wave functions with spin
indices carrying representations of the little algebras. If we want to
describe interactions where energy and momentum can be transfered from
one particle onto another, we need to leave the mass shell. And we
need a way to describe several particle types and their coupling in
one common formalism.

This can be done by introducing Lorentz spinor wave functions. That
is, tensor products of the algebra of functions on spacetime with a
finite vector space containing the spin degrees of freedom, the whole
space carrying a tensor representation of the Lorentz algebra. The
additional mathematical structure we need in order to couple two wave
functions is provided by the multiplication within the algebra of
space functions. However, these Lo\-rentz spinor representations are
not irreducible. Therefore, only an irreducible subrepresentation can
be the space of physical states. This subrepresentation is
conveniently described as kernel of a linear operator $\mathbb{A}$,
that is, we demand all physical states $\psi$ to satisfy the wave
equation $\mathbb{A}\psi = 0$.

This line of thought relies on the sole assumption that the
Poincar{\'e} algebra describes the basic symmetry of spacetime. In
this work we replace the Poincar\'e algebra by its $q$-deformation
\cite{Ogievetskii:1992a}, which describes the basic symmetry of
$q$-deformed spacetime. Then we construct $q$-wave equations
proceeding in exactly the same way as in the undeformed case. 

In \cite{Dobrev:1994} similar covariance arguments have been used to
find wave equations for a certain deformation of the conformal
symmetry algebra, which, however, does not contain the $q$-Poincar\'e
algebra we consider here. In order to construct $q$-deformed
relativistic wave equations various other methods have been proposed,
based on $q$-Clifford algebras \cite{Schirrmacher:1992}, $q$-deformed
co-spinors \cite{Pillin:1994b}, or differential calculi on quantum
spaces \cite{Song:1992,Meyer:1995,Podles:1996}, leading to mutually
different results. While each approach may be justified in its own
right, the situation as a whole is unsatisfactory since it should be
possible to determine the wave equations \emph{uniquely} as in the
undeformed case \cite{BarutRaczka} without needing any additional
mathematical structure besides the $q$-Poincar\'e algebra and the
basic apparatus of quantum mechanics.

Throughout this article, it is assumed that $q$ is a real number
$q>1$. We will frequently use the abbreviations $\lambda = q - q^{-1}$
and $[2] = q+q^{-1}$. The lower case Greek letters $\mu$, $\nu$,
$\sigma$, $\tau$ denote 4-vector indices running through
$\{0,-,+,3\}$. Lower 4-indices are raised by $P^\mu := \eta^{\mu\nu}
P_\nu$ with the 4-metric $\eta^{\mu\nu}$ of Eq.~\eqref{eq:FourMetric}
such that $S_\mu P^\mu$ is a scalar.  The upper case Roman letters
$A$, $B$, $C$ denote 3-vector indices running through $\{-1,0,+1\} =
\{-,3,+\}$.  A very short introduction to the $q$-Poincar{\'e} algebra
is given in Appendix~\ref{sec:AppPoin}. Some more mathematical
background information for this article has been compiled in
\cite{Blohmann}.

\section{\textit{q}-Spinor Wave Functions}

\subsection{General \textit{q}-Wave Equations}

We seek linear wave equations $\mathbb{A}\psi = 0$, where $\mathbb{A}$
is a linear operator. For $\ker \mathbb{A}$ to be a subrepresentation,
the operator must satisfy
\begin{equation}
\label{eq:WaveCondition1}
  \mathbb{A}\psi = 0 \quad\Rightarrow\quad \mathbb{A}h\psi = 0
\end{equation}
for all $q$-Poincar{\'e} transformations $h$. Depending on the particle
type under consideration we might include charge and parity
transformations. $\mathbb{A}$ is not unique since the wave
equations for $\mathbb{A}$ and $\mathbb{A}'$ must be considered
equivalent as long as their solutions are the same, $\ker(\mathbb{A})
= \ker(\mathbb{A}')$. 

Ideally, $\mathbb{A}$ is a projection operator, $\mathbb{A}=\Proj$,
with $\Proj^2=\Proj$, $\Proj^* = \Proj$.
Condition~\eqref{eq:WaveCondition1} is then equivalent to
\begin{equation}
\label{eq:WaveCondition2}
  [\Proj,h] = 0
\end{equation}
for all $q$-Poincar{\'e} transformations $h$. Whether the wave equation is
written with a projection is a matter of convenience. The Dirac
equation is commonly written with such a projection which is determined
uniquely (up to complement) by condition~\eqref{eq:WaveCondition2}.
For the Maxwell equations a projection can be found but yields a second
order differential equation. For this reason, the Maxwell equations
are commonly described by a more general operator $\mathbb{A}$, which
leads to a first order equation.

\subsection{\textit{q}-Lorentz Spinors}

We define a general, single particle $q$-Lorentz spinor wave function
as element of the tensor product $\Scal\otimes \Xcal$ of a finite
vector space $\Scal = \mathbb{C}^n$ holding the spin degrees of
freedom and the space of $q$-Minkowski space functions $\Xcal = \Mink$
(App.~\ref{sec:AppPoin}).

Let $\{ e_i \}$ be a basis of $\Scal$ transforming under a $q$-Lorentz
transformation $h\in\Hcal=\slC$ as $h\tr e_j = e_i \,\rho(h)^i{}_j$,
where $\rho : \Hcal \rightarrow \mathrm{End}(\Scal)$ is the
representation map. Any spinor $\psi \in \Scal \otimes \Xcal$ can be
written as
\begin{equation}
  \psi = e_j \otimes \psi^j \,,
\end{equation}
where $j$ is summed over and the $\psi^j$ are elements of $\Xcal$. The
action of $h\in\Hcal$ on a spinor is the tensor action
\begin{equation}
  h\psi = (h_{(1)}\tr e_j) \otimes (h_{(2)}\tr \psi^j)
  = e_i\otimes \rho(h_{(1)})^i{}_j  (h_{(2)}\tr \psi^j) \,.
\end{equation}
This tells us that the $\Xcal$-valued components of $h\psi$ are given
by
\begin{equation}
\label{eq:Wave1}
  (h\psi)^i = \rho(h_{(1)})^i{}_j  (h_{(2)}\tr \psi^j) \,.
\end{equation}
The transformation of $\psi$ can easily be generalized to the case
where $\Scal$ carries a tensor representation of two finite
representations. For the components of spinors with two indices we
would get
\begin{equation}
\label{Wave2}
  (h\psi)^{ij} = \rho(h_{(1)})^i{}_{i'}
               \rho'(h_{(2)})^j{}_{j'}  (h_{(3)}\tr \psi^{i'j'}) \,,
\end{equation}
where $\rho$ and $\rho'$ are the representation maps of the first and
second index, respectively. 

Furthermore, we get spinors from the action of tensor operators. Let
$T^i$ be a $\rho$-tensor operator with respect to the left Hopf
adjoint action, $\adL h \tr T^i \equiv h_{(1)} T^i S(h_{(2)}) =
\rho(Sh)^i{}_j T^j$, and $\psi = e_j \otimes \psi^j$ a $\rho'$-spinor.
Let us define the components of a spinor $\phi^{ij}$ with two indices
by
\begin{equation}
\label{eq:Bong}
  \phi^{ij} := (T^i \psi)^j \,.
\end{equation}
How does this new array of wave functions $\phi^{ij}$ transform under
$q$-Lorentz transformations? Letting $h$ act from the left, we find
\begin{equation}
  h\phi^{ij} = \rho(h_{(1)})^j{}_{j'}(h_{(2)}\tr \phi^{ij'}) \,,
\end{equation}
that is, $h$ acts only on the index of the wave functions $\psi^j$.
However, if we transform $\phi^{ij}$ by transforming $\psi$ inside
Eq.~\eqref{eq:Bong},
\begin{align}
  \bigl(T^i (h\psi)\bigr)^j &= \bigl((T^i h)\psi\bigr)^j 
  = \bigl( h_{(2)}  [ \adL S^{-1}(h_{(1)}) \tr T^i ] 
    \psi \bigr)^{j} \notag\\ 
  &= \bigl( \rho(h_{(1)})^i{}_{i'}  h_{(2)}T^{i'} \psi \bigr)^{j} 
  = \rho(h_{(1)})^i{}_{i'}  h_{(2)} \phi^{i'j} \notag \\
  &= \rho(h_{(1)})^i{}_{i'} \rho'(h_{(2)})^j{}_{j'} 
     (h_{(3)} \tr \phi^{i'j'})  \,,
\end{align}
we find that $\phi^{ij}$ transforms as a $\rho\otimes\rho'$-spinor.
Note, that for the last calculation the order in the tensor product
$\Scal\otimes \Xcal$ is essential. It would not have worked out as
nicely if we had constructed the spinor space as $\Xcal\otimes\Scal$.
Chief examples of this construction would be the gauge term
$P^\mu\phi$ of the vector potential $A^\mu$, or the derivatives of the
vector potential $P^\mu A^\nu$ used to construct the electromagnetic
field strength tensor $F^{\mu\nu}$.

\subsection{\textit{q}-Derivatives}

The $q$-Poincar\'e algebra is the Hopf semidirect product of the
$q$-Lorentz algebra and the $q$-Minkowski algebra generated by the
momentum 4-vector $P_\mu$ (Appendix). It becomes a Hopf algebra upon
bosonisation \cite{Majid:1993}. Within the braided tensor product, the
coproduct of the momenta takes the natural form $\Delta(P^\mu) = P^\mu
\underline{\otimes} 1 + 1 \underline{\otimes} P^\mu$ or, explicitly,
\begin{equation}
\label{eq:Pcoproduct}
\begin{split}
  \Delta(P^\mu) &:=
  P^\mu \otimes 1 + \R^{-1}( \kappa \otimes P^\mu)\R\\
  &= P^\mu \otimes 1 + \kappa L_+^\mu{}_\nu \otimes P^\nu\,,
\end{split} 
\end{equation}
where $\R = \R^{(1)} \otimes \R^{(2)}$ is the real universal
$\R$-matrix of the $q$-Lorentz algebra, $\kappa$ is a group-like
scaling operator, $P^\mu \kappa = q P^\mu \kappa$, which commutes with
all Lorentz generators, and the $L$-matrix $L_+^\mu{}_\nu := \R^{(1)}
\Lambda(\R^{(2)})^\mu{}_\nu$ is given by the 4-vector representation
of the second tensor factor of $\R$.

As in the undeformed case, we can assume that the spinorial degrees of
freedom carry the trivial representation of the momentum algebra, that
is, $P_\mu \tr e_j = \varepsilon(P_\mu) e_j = 0$ and $\kappa \tr e_j =
\varepsilon(\kappa) e_j = e_j$. Hence, $P_\mu$ acts on a $\rho$-spinor
as $P_\mu \psi = \rho(L_+^\mu{}_\nu)e_j \otimes (P^\nu \tr \psi^j)$,
that is,
\begin{equation}
\label{eq:putschi1}
  (P^\mu \psi)^i =
  \rho(L_+^{\mu}{}_{\nu})^i{}_j \,(P^{\nu} \tr \psi^j) \,.
\end{equation}
This equation yields a well defined action for both, the antireal and
the real $\R$-matrix. For the following computations we will chose the
antireal $\R$-matrix, for only then the action of the momenta is
compatible with the $*$-structure.  The $L$-matrix for this case has
been calculated in \cite{Blohmann}. The results for the alternative
choice of the real $\R$-matrix are given in the Appendix.

The action of $P^\mu$ on each component $\psi^j \in \Xcal$ can be
viewed as derivation within the algebra of $q$-Minkowski space
functions $\Xcal$,
\begin{equation}
  \partial^\mu := 1\otimes \I P^\mu \,.
\end{equation}
Now we can interpret an operator linear in the momenta as
$q$-differential operator. If $C_\mu = C_\mu \otimes 1$ are operators
that act on the spinor indices only,
\begin{equation}
  \I\,C_\mu  P^\mu = C_\mu\,
  \rho(L_+^{\mu}{}_{\nu}) \partial^{\nu}
  = \tilde{C}_{\nu} \partial^{\nu} \,,
\end{equation} 
where
\begin{equation}
\label{eq:Optilde}
  \tilde{C}_{\nu} := C_\mu\,\rho(L_+^{\mu}{}_{\nu})
\end{equation}
such that $\tilde{C}_{\nu}$ still acts on the spinor index only, while
$\partial^{\nu}$ acts componentwise, so the two operators commute
$[\tilde{C}_{\mu},\partial^{\nu}] = 0$. We will calculate the
transformation $C_\mu \rightarrow \tilde{C}_\mu$ for particular
representations below. Finally, we remark that for the mass Casimir we
have within the spinor representation $P_\mu P^\mu = \R^{-1}(1 \otimes
P_\mu P^\mu)\R = 1 \otimes P_\mu P^\mu$, hence, $P_\mu P^\mu = -
\partial_\mu \partial^\mu$. This means, that mass irreducibility for a
spinor is the same as mass irreducibility for each component of the
spinor.

\subsection{Conjugate Spinors}

One of the effects of using Lorentz spinors is that the underlying
representations can no longer be unitary, since there are no unitary
finite representations of the non-compact Lorentz algebra --- in the
$q$-deformed as well as in undeformed case. However, we can introduce
non-degenerate but indefinite bilinear forms playing the role of the
scalar product. With respect to these pseudo scalar products the
spinors carry $*$-representations, that is, the $*$-operation on the
algebra side is the same as the pseudo adjoint on the operator side.

The problem of non-unitarity arises from the finiteness of the spin
part $\Scal$ within the space of spinor wave functions $\Scal\otimes
\Xcal$, so we can assume that the wave function part $\Xcal$ does
carry a $*$-representation. It is then sufficient to redefine the
scalar product on $\Scal$ only. Consider a $D^{(j,0)}$ representation
of $\slC$ with orthonormal basis $\{e_n\}$.  We want to define a
pseudo scalar product by $( e_m | e_n ) := A_{mn}$ such that
\begin{equation}
\label{eq:pseudounitary}
  ( e_m | (g\otimes h)\tr e_n ) = ((g\otimes h)^*\tr e_m | e_n )
\end{equation}
for any $g\otimes h \in \slq\otimes\slq \cong \slC$. For a pseudo
scalar product we must suppose $A_{mn}$ to be a non-degenerate,
hermitian, but not necessarily positive definite matrix. Inserting the
definition of the pseudo scalar product, the pseudo-unitarity
condition~\eqref{eq:pseudounitary} reads
\begin{equation}
\label{eq:pseudounitary2}
\begin{split}
  ( e_m | (g\otimes h) \tr e_n )
  &= ( e_m |\,e_{n'} \rho^j(g)^{n'}{}_{n}\,\varepsilon(h) )\\
  &= A_{mn'} \rho^j(g)^{n'}{}_{n}\,\varepsilon(h) \\
  &{}\stackrel{!}{=} ( (g\otimes h)^* \tr e_m |\, e_n )\\
  &= ( e_{m'}\varepsilon(g^*) \rho^j(h^*)^{m'}{}_{m} |\, e_{n} ) \\
  &= A_{m'n} \,\overline{\varepsilon(g^*)\rho^j(h^*)^{m'}{}_{m} } \\
  &=  A_{m'n} \,\varepsilon(g)\rho^j(h)^{m}{}_{m'} \,,
\end{split}
\end{equation}
where we have used that $(g\otimes h)^* = \R_{21}(h^* \otimes
g^*)\R_{21}^{-1}$ \cite{Majid} and $\varepsilon(\R_{[1]})\R_{[2]}=1$.
Traditionally, the pseudo scalar product is not described by a matrix
$A_{mn}$ but by introducing a conjugate spinor basis $\{\bar{e}_n\}$
by
\begin{equation}
  \bar{e}_m := e_{m'} A_{m'm} \,.
\end{equation}
Using~\eqref{eq:pseudounitary2} the conjugate basis turns out to
transform as
\begin{equation}
\begin{split}
  (g\otimes h) \tr \bar{e}_n 
  &= e_{m'} \rho^j(g)^{m'}{}_{m} \varepsilon(h) \,A_{mn}\\
  &=e_{m'} A_{m'n'} \varepsilon(g)\rho^j(h)^{n'}{}_{n} \\
  &= \bar{e}_{n'} \varepsilon(g) \rho^j(h)^{n'}{}_{n} \,,
\end{split}
\end{equation}
that is, $\bar{e}_n$ must transform according to a
$D^{(0,j)}$ representation. $D^{(j,0)}$ and $D^{(0,j)}$ being
inequivalent representations, the conjugate basis $\bar{e}_n$ cannot
be expressed as a linear combination of the original basis vectors
$e_n$. In order to allow for a conjugate spinor basis we must consider
a representation which contains both, $D^{(j,0)}$ and $D^{(0,j)}$, and
thus at least their direct sum $D^{(j,0)}\oplus D^{(0,j)}$ as
subrepresentation.

So far, it seems that everything is almost trivially analogous to the
undeformed case. It is not. If we consider irreducible representations
of mixed chirality, $D^{(i,j)}$, we find that the appearance of the
$\R$-matrix in $(g\otimes h)^*$ makes it impossible
to define conjugate spinors. It only works for $D^{(j,0)}$, because
$\rho^0 = \varepsilon$ and $\varepsilon(\R_{[1]})\R_{[2]} = 1$.
Fortunately, we do have conjugate spinors for the most interesting
cases: Dirac spinors ($D^{(\frac{1}{2},0)}\oplus D^{(0,\frac{1}{2})}$)
and the Maxwell tensor ($D^{(1,0)}\oplus D^{(0,1)}$). For these cases
everything is analogous to the undeformed case.

Let us consider a $D^{(j,0)}\oplus D^{(0,j)}$ representation with
basis $\{e^\mathrm{L}_n\}$ for the left chiral subrepresentation
$D^{(j,0)}$ and the basis $\{e^\mathrm{R}_n\}$ for $D^{(0,j)}$. We
define the conjugate basis by $\overline{e^\mathrm{L}_n} :=
e^\mathrm{R}_n$ and $\overline{e^\mathrm{R}_n} = e^\mathrm{L}_n$. Let
us call $\mathcal{P}$ the parity operator that exchanges the left and
right chiral part. Its matrix representation with respect to the
basis $\{e^\mathrm{L}_n , e^\mathrm{R}_n\}$ is
\begin{equation}
\label{eq:Parity}
  \mathcal{P}_{mn}=
  \begin{pmatrix} 0 & 1 \\ 1 & 0 \end{pmatrix}\,,
\end{equation}
where $1$ is the $(2j+1)$-dimensional unit matrix. This is the matrix
that represents our new pseudo scalar product as a bilinear form.  The
pseudo Hermitian conjugate of some operator $A$ can now be written as
\begin{equation}
  j(A) := \mathcal{P}A^\dagger \mathcal{P} \,,
\end{equation}
which is an involution because $\mathcal{P}^\dagger =
\mathcal{P}^{-1}$ and an algebra anti-homomorphism because
$\mathcal{P} = \mathcal{P}^{-1}$.

We apply this result to the whole space of spinor wave functions
$\Scal\otimes \Xcal$. Let us assume that the scalar product of two
wave functions $f,g\in\Xcal$ can be written (at least formally) as
some sort of integral $\lrAngle{f | g } = \int f^* g$.  The pseudo
scalar product of two $D^{(j,0)}\oplus D^{(0,j)}$ spinors $\psi$,
$\phi$ becomes
\begin{equation}
\begin{split}
  (\psi|\phi) &= (e_m\otimes \psi^m|e_n \otimes \phi^n)
  =(e_m|e_n) \lrAngle{ \psi^m|\phi^n} \\
  &= {\textstyle\int} (\psi^m)^* \mathcal{P}_{mn} \phi^n
  = {\textstyle\int} \bar{\psi}^n \phi^n \,,
\end{split}
\end{equation}
with the conjugate spinor wave function defined as
\begin{equation}
\label{eq:Wave3}
  \bar{\psi}^n := (\psi^m)^* \mathcal{P}_{mn} \,.
\end{equation}
In summary, we have convinced ourselves that in the case of
$D^{(j,0)}\oplus D^{(0,j)}$ representations the conjugation of
spinors, of spinor wave functions, and of operators works exactly as
in the undeformed case. 


\section{The \textit{q}-Dirac Equation}
\label{sec:MainContrib4b}

\subsection{The \textit{q}-Dirac Equation for the Rest States}

In this section we consider $q$-Dirac spinors $\psi = e_i
\otimes\psi^i$ with the spin part transforming according to a
$D^{(\frac{1}{2},0)}\oplus D^{(0,\frac{1}{2})}$ representation
\cite{Blohmann}. We want to write the $q$-Dirac equation as expression
which involves momenta only to first order, corresponding to a first
order differential equation 
\begin{equation}
\label{eq:qDirac}
  \Proj \psi := \frac{1}{2m}(m + \gamma_\mu P^\mu)\psi = 0 \,,
\end{equation}
with the $\gamma_\mu$ being some operators acting on $\psi^i$. We can
already say that $\gamma_\mu$ must be a left 4-vector operator.  If it
were not, $\gamma_\mu P^\mu$ would not be scalar and, hence, would not
commute with the $q$-Lorentz transformations as required in
Eq.~\eqref{eq:WaveCondition2}. 

We consider here a massive $q$-Dirac spinor representation, so there
is a set of rest states $\psi$ which the momenta act upon as $P^0\psi
= m \psi$, $P^A\psi = 0$ \cite{Blohmann:2001a}. We start the search
for a projection $\Proj$ that reduces the $q$-Dirac representation by
computing how it acts on these rest states, where we have
\begin{equation}
  \Proj_0 = \tfrac{1}{2}(1 + \gamma_0) \,,
\end{equation}
the zero indicating that $\Proj_0$ acts on the rest states only. We
assume that we can realize the operator $\gamma_0$ as $4\times
4$-matrix that acts on the spin degrees of freedom only. This is not
unreasonable, for if $\gamma_\mu$ is a set of matrices that form a
$4$-vector operator in the $D^{(\frac{1}{2},0)}\oplus
D^{(0,\frac{1}{2})}$ representation then $\gamma_\mu \otimes 1$ will
also be a $4$-vector operator in the representation of spinor wave
functions. So let us assume we can write $\Proj_0 = \Proj_0
\otimes 1$ in block form as $\Proj_0 = (\begin{smallmatrix} A & B \\
  C & D \end{smallmatrix})$, where $A$, $B$, $C$, $D$ are $2\times
2$-matrices.

The restriction of condition~\eqref{eq:WaveCondition2} to the rest
states means that $\Proj_0$ must commute with the little algebra. The
little algebra for the massive case is the $\suq$ subalgebra of
rotations \cite{Blohmann:2001a}. A rotation $l\in\suq$ is represented
by
\begin{equation}
  \rho(l) = \begin{pmatrix}
    \rho^{\frac{1}{2}}(l) & 0 \\ 0 & \rho^{\frac{1}{2}}(l)
  \end{pmatrix} \,.
\end{equation} 
Since the $\rho^{\frac{1}{2}}$ representations of the rotations
generate all $2\times2$-matrices (the $q$-Pauli matrices are a basis),
$\Proj_0$ will only commute with all rotations if $A$, $B$, $C$, $D$
are numbers, that is, complex multiples of the unit matrix.

Furthermore, $\Proj_0$ has to be a projection operator, $\Proj_0^2 =
\Proj_0$, $\Proj_0^\dagger = \Proj_0$, and, as in the undeformed case,
we require it to commute with the parity operator,
$[\Proj_0,\mathcal{P}]=0$.  Together these conditions fix $\Proj_0$
and hence $\gamma_0$ uniquely to
\begin{equation}
  \gamma_0 = \begin{pmatrix} 0 & 1 \\ 1 & 0 \end{pmatrix} \,,
\end{equation}
the same as in the undeformed case.

\subsection{The \textit{q}-Gamma Matrices and the \textit{q}-Clifford Algebra}
\label{sec:gammaboost}

If $\gamma_0$ is to be a 4-vector operator, we have to define the
other gamma matrices as in Eq.~\eqref{eq:Boost1} by
\begin{equation}
\label{eq:boostgamma}
\begin{aligned}
 \gamma_- &= \adL(-q^{-\frac{1}{2}}\lambda^{-1}[2]^{\frac{1}{2}} \,c)
             \tr \gamma_0 \\ 
 \gamma_+ &= \adL(q^{\frac{1}{2}}\lambda^{-1}[2]^{\frac{1}{2}} \,b)
             \tr \gamma_0 \\ 
 \gamma_3 &= \adL(\lambda^{-1}\,(d-a)) \tr \gamma_0 \,,
\end{aligned}
\end{equation}
where the adjoint action is understood with respect to the $q$-Dirac
representation. In order to compute this explicitly, we have to
calculate the representations of the boosts~\eqref{eq:boostdef} first.
\begin{subequations}
\begin{align}
  \rho(a) &=  \begin{pmatrix}
    \rho^{\frac{1}{2}}(K^{\frac{1}{2}}) & 0 \\
    0 & \rho^{\frac{1}{2}}(K^{-\frac{1}{2}}) \end{pmatrix}\\
  \rho(b) &=  \begin{pmatrix} 0 & 0 \\
    0 & q^{-\frac{1}{2}}\lambda \rho^{\frac{1}{2}}(K^{-\frac{1}{2}}E)
  \end{pmatrix}\\
  \rho(c) &=  \begin{pmatrix} 
    -q^{\frac{1}{2}}\lambda \rho^{\frac{1}{2}}(FK^{\frac{1}{2}})
    & 0 \\ 0 & 0 \end{pmatrix}\\
  \rho(d) &=  \begin{pmatrix}
    \rho^{\frac{1}{2}}(K^{-\frac{1}{2}}) & 0 \\
    0 & \rho^{\frac{1}{2}}(K^{\frac{1}{2}}) \end{pmatrix}
\end{align}
\end{subequations}
This gives us for example
\begin{equation}
\begin{split}
  \gamma_+
  &= \adL(q^{\frac{1}{2}}\lambda^{-1}[2]^{\frac{1}{2}}\,b)\tr\gamma_0 \\
  &= q^{\frac{1}{2}}\lambda^{-1}[2]^{\frac{1}{2}}
    [\rho(b)\gamma_0\rho(a) - q \rho(a)\gamma_0\rho(b) ] \\
  &= [2]^{\frac{1}{2}}\left[
    \begin{pmatrix} 0 & 0 \\
      \rho^{\frac{1}{2}} ( K^{-\frac{1}{2}} E K^{\frac{1}{2}} ) & 0
    \end{pmatrix} 
    -q \begin{pmatrix}
      0 & \rho^{\frac{1}{2}}(E) \\ 0 & 0
    \end{pmatrix} \right] \\ 
  &= \begin{pmatrix} 0 & q \,\sigma_+ \\
    -q^{-1}\sigma_+ & 0 \end{pmatrix}, 
\end{split}
\end{equation}
where $\sigma_+$ is one of the $q$-Pauli matrices defined as
$q$-Clebsch-Gordan coefficients or, equivalently, as
spin-$\tfrac{1}{2}$ representation of the angular momentum generators,
$\sigma\!_A = [2]\rho^{\frac{1}{2}}(J_A)$ \cite{Blohmann}. The
analogous calculations for $\gamma_-$ and $\gamma_+$ yield
\begin{equation}
\label{eq:gamma1}
  \gamma_0 = \begin{pmatrix} 0 & 1 \\ 1 & 0 \end{pmatrix}, \qquad
  \gamma_A = \begin{pmatrix} 0 & q\, \sigma\!_A \\
      -q^{-1}\sigma\!_A & 0 \end{pmatrix} \,,
\end{equation}
where $A$ runs as usual through $\{-,+,3\}$.

This result can be easily generalized to higher spin. All we have to
do for a massive $D^{(j,0)}\oplus D^{(0,j)}$-spinor is to replace
$\rho^{\frac{1}{2}}$ with $\rho^j$. The result is higher dimensional
$\gamma$-matrices
\begin{equation}
\label{eq:GammaGeneral}
  \gamma_0^{(j)} = \begin{pmatrix} 0 & 1 \\ 1 & 0 \end{pmatrix},\,
  \gamma_A^{(j)} = [2] \begin{pmatrix} 0 & q\,\rho^j(J_A)  \\
    -q^{-1}\rho^j(J_A) & 0 \end{pmatrix} \,.
\end{equation}
If we want to write the $q$-Dirac equation as $q$-differential
equation, we need to calculate $\tilde{\gamma}_\mu$ by
formula~\eqref{eq:Optilde}. For the $q$-Pauli matrices we get
\begin{subequations}
\begin{align}
  \sigma\!_A\,\rho^{(\frac{1}{2},0)}
  \bigl((L^\Lambda_{\mathrm{I}+})^{A}{}_{B}\bigr)
  &= q^2\tilde{\sigma}\!_B \\ 
  \sigma\!_A\,\rho^{(0,\frac{1}{2})}
  \bigl((L^\Lambda_{\mathrm{I}+})^{A}{}_{B}\bigr)
  &= q^{-2}\tilde{\sigma}_B \,,
\end{align}
\end{subequations}
where we have used a variant of the $q$-Pauli matrices, defined as
$\tilde{\sigma}\!_A := - [2] \rho^{\frac{1}{2}}(SJ_A)$. Explicitly,
these are
\begin{equation*}
  \tilde{\sigma}_- = {[2]}^{\frac{1}{2}}
    \begin{pmatrix}
      0 & q^{\frac{1}{2}} \\ 0 & 0
    \end{pmatrix}, 
  \tilde{\sigma}_+ = [2]^{\frac{1}{2}}
    \begin{pmatrix}
      0 & 0 \\ -q^{-\frac{1}{2}} & 0
    \end{pmatrix},
  \tilde{\sigma}_3 =
    \begin{pmatrix} -q^{-1} & 0 \\ 0 & q \end{pmatrix} 
\end{equation*}
with respect to the $\{-,+\}$ basis. For the transformed $q$-gamma
matrices we thus obtain
\begin{equation}
\label{eq:gamma2}
  \tilde{\gamma}_0 =
    \begin{pmatrix} 0 & 1 \\ 1 & 0 \end{pmatrix}\,, \qquad
  \tilde{\gamma}_A =
    \begin{pmatrix} 0 & q^{-1}\, \tilde{\sigma}\!_A \\
      -q \tilde{\sigma}\!_A & 0 \end{pmatrix} \,, 
\end{equation}
and the $q$-Dirac equation written as $q$-differential equation becomes
\begin{equation}
  (m - \I \tilde{\gamma}_\mu \partial^\mu)\psi = 0 \,.
\end{equation}
After lengthy calculations we find that the gamma matrices satisfy the
relations
\begin{equation}
\label{eq:Clifford}
  \tilde{\gamma}_\sigma  \tilde{\gamma}_\tau
  = \eta_{\tau\sigma} + \tilde{\gamma}_\mu \tilde{\gamma}_\nu
  \Proj_\mathrm{A}^{\nu\mu}{}_{\tau\sigma}
  \quad \Leftrightarrow \quad
  \tilde{\gamma}_\mu \tilde{\gamma}_\nu
  \Proj_\mathrm{S}^{\nu\mu}{}_{\sigma\tau} = \eta_{\sigma\tau}\,,
\end{equation}
where $\Proj_\mathrm{A}$ is the $q$-antisymmetrizer and
$\Proj_\mathrm{S} = 1 - \Proj_\mathrm{A}$ is the $q$-symmetrizer of
the Clebsch-Gordan series~\eqref{eq:CGseries}. This is the
$q$-deformation of the Clifford algebra relations, from which now
follows that the square of $q$-Dirac operator is indeed the mass
Casimir,
\begin{equation}
  (\tilde{\gamma}_\mu \partial^\mu)^2 = \partial_\mu \partial^\mu
  = -P_\mu P^\mu \,.
\end{equation}
As in the undeformed case we conclude that a solution $\psi$ to the
$q$-Dirac equation satisfies automatically the mass shell condition
$P_\mu P^\mu \psi = m^2 \psi$, and that $\Proj =
\frac{1}{2m}(m+\gamma_\mu P^\mu)$ really is a projection operator.

One could have started directly from relations~\eqref{eq:Clifford}
trying to find matrices that satisfy them \cite{Schirrmacher:1992}.
This approach has a number of disadvantages: a) It is computationally
much more cumbersome than boosting $\gamma_0$. b) The result is not
unique, that is, we would get many solutions to the $q$-Clifford
algebra not knowing which representations they belong to. c) Having
determined a solution $\tilde{\gamma}_\mu$, the covariance of the
$q$-Dirac equation remains unclear as $\tilde{\gamma}_\mu$ cannot be a
4-vector operator.

\subsection{The Zero Mass Limit and the \textit{q}-Weyl Equations}

The zero mass limit of the $q$-Dirac equation is formally
\begin{equation}
   \mathbb{A}\psi := \gamma_\mu P^\mu \psi = 0 \,,
\end{equation}
where $\mathbb{A}$ is no longer a projection operator.  The wave
equation is now decoupled into two independent equations for a left
handed $D^{(\frac{1}{2},0)}$-spinor $\psi_\mathrm{L}$ and a right
handed $D^{(0,\frac{1}{2})}$-spinor $\psi_\mathrm{R}$,
\begin{xalignat}{2}
  \sigma\!_A P^A \psi_\mathrm{L} &= q^{-1}P^0 \psi_\mathrm{L} \,,&
  \sigma\!_A P^A \psi_\mathrm{R} &= -q P^0 \psi_\mathrm{R} \,,
\end{xalignat}
the $q$-Weyl equations for massless left and right handed
spin-$\frac{1}{2}$ particles. Written as $q$-differential equation
they become
\begin{xalignat}{2}
  \tilde{\sigma}\!_A \partial^A \psi_\mathrm{L}
  &= -q \partial^0 \psi_\mathrm{L} \,,& 
  \tilde{\sigma}\!_A \partial^A \psi_\mathrm{R}
  &= q^{-1} \partial^0 \psi_\mathrm{R} \,.
\end{xalignat}
The operator $\mathbb{A}$ inherits property~\eqref{eq:WaveCondition1}
from $\Proj$, so $\mathbb{A}\psi = 0$ is a viable wave equation. On
the massless momentum eigenspace \cite{Blohmann:2001a} where
$(P_0,P_-,P_+,P_3) = (k,0,0,k)$, $\mathbb{A}$ acts
as $ \mathbb{A}_0 = k (\begin{smallmatrix} 0 & 1-q\sigma_3 \\
  1+q^{-1}\sigma_3 & 0 \end{smallmatrix})$. The kernel of this
operator is 2-dimensional, the solution states corresponding to
helicity $\pm\frac{1}{2}$.

If we generalize these considerations to higher spin $D^{(j,0)}\otimes
D^{(0,j)}$ Dirac type spinors, we find that the corresponding operator
$\mathbb{A}$ has zero kernel, so the space of solutions is trivial.
This applies in particular to $q$-Maxwell spinors. Therefore, we need
a different approach to find the $q$-Maxwell equations --- in complete
analogy to the undeformed case \cite{BarutRaczka}.

\section{The \textit{q}-Maxwell Equations}
\label{sec:MainContrib4c}

\subsection{The \textit{q}-Maxwell Equations in the Momentum
  Ei\-gen\-spa\-ces}

In this section we consider massless $D^{(1,0)}\oplus D^{(0,1)}$
spinors. According to the Clebsch-Gordan series~\eqref{eq:CGseries}
this type of spinor is equivalent to a $q$-antisymmetric tensor
$F^{\mu\nu}$ with two 4-vector indices. These are the types of spinors
commonly used to describe the electromagnetic field.

We start our calculations in the massless momentum eigenspace with
momentum eigenvalues $(P_0,P_-,P_+,P_3) = (k,0,0,k)$ for some real
parameter $k$. It has been shown in \cite{Blohmann:2001a} that this
eigenspace is invariant under the little algebra generated by the
group-like generator of $q$-rotations around the $z$-axis $K$ and
\begin{equation}
\label{eq:Little3}
  N_- := q^{\frac{1}{2}}[2]^{\frac{1}{2}} ac ,\,\,
  N_+ := q^{\frac{1}{2}}[2]^{\frac{1}{2}} bd ,\,\,
  N_3 := 1 +[2] bc \,.
\end{equation}
The irreducible $*$-representations of this algebra are
one-dimensional, given by $K = \kappa = N_3$ and $N_\pm = 0$ for real
$\kappa$. Within the momentum eigenspace the little algebra acts only
on the spinor index, here, by the $D^{(1,0)}\oplus D^{(0,1)}$ matrix
representation
\begin{equation}
\label{eq:Maxwell1}
\begin{gathered}
  K =
    \begin{pmatrix}\rho^1(K)&0\\0 &\rho^1(K) \end{pmatrix},\quad
  N_- = -q[2]
    \begin{pmatrix} \rho^1(J_-) & 0 \\ 0 & 0 \end{pmatrix}\\
  N_+ = -q^{-1}[2]
    \begin{pmatrix} 0 & 0 \\ 0 & \rho^1(J_+)  \end{pmatrix}, \quad
  N_3 =
    \begin{pmatrix} 1 & 0 \\ 0 & 1 \end{pmatrix}
\end{gathered}
\end{equation}
where $\rho^1$ is the vector representation of $\suq$.  We seek an
operator $\Proj_0 = \Proj_0 \otimes 1$ that projects onto an
irreducible subrepresentation of the momentum eigenspace. As before,
we write it in block form as
$\Proj_0 = (\begin{smallmatrix} A & B \\
  C & D \end{smallmatrix})$, where $A$, $B$, $C$, $D$ are $3\times
3$-matrices. We must have $\Proj_0^\dagger =\Proj_0$, so $A$ and $D$
must be Hermitian matrices and $C=B^\dagger$.  Within an irreducible
representation of the little algebra we have $N_\pm = 0$, so we must
demand $N_\pm \Proj_0 = 0$. This leads to the conditions
\begin{equation}
\begin{gathered}
   \rho^1(J_-)\,A = 0\,,\qquad  \rho^1(J_+)\,D = 0 \\
   \rho^1(J_-)\,B = 0\,,\qquad  \rho^1(J_+)\,B^\dagger = 0 \,.
\end{gathered}
\end{equation}
To satisfy these conditions $A$, $B$, and $D$ must be of the form
\begin{xalignat}{3}
  A &= \begin{pmatrix}
    \alpha & 0 & 0 \\
    0 & 0 & 0 \\
    0 & 0 & 0  \end{pmatrix}, &
  B &= \begin{pmatrix}
    0 & 0 & \beta \\
    0 & 0 & 0 \\
    0 & 0 & 0  \end{pmatrix}, &
  D &= \begin{pmatrix}
    0 & 0 & 0 \\
    0 & 0 & 0 \\
    0 & 0 & \delta  \end{pmatrix},
\end{xalignat}
for $\alpha$, $\delta$ real and $\beta$ complex. Furthermore,
$\Proj_0$ must project onto an eigenvector of $K$. From this it
follows that $\beta = 0$ and either $\alpha=1$, $\delta=0$ or
$\alpha=0$, $\delta=1$. To summarize, there are two possible
projections
\begin{xalignat}{2}
  \Proj_\mathrm{L} &= \begin{pmatrix}
    1 &   &   &   \\
      & 0 &   &   \\
      &   & \diagdown  &   \\
      &   &   & 0 \end{pmatrix}, &
  \Proj_\mathrm{R} &= \begin{pmatrix}
    0 &   &   &   \\
      & \diagdown &   &   \\
      &   & 0 &   \\
      &   &   & 1 \end{pmatrix} 
\end{xalignat}
projecting each on a irreducible one-dimensional representation of the
little algebra. The image of $\Proj_\mathrm{L}$ is part of the left
handed $D^{(1,0)}$ component while $\Proj_\mathrm{R}$ projects onto
the right handed $D^{(0,1)}$ component of the spinor. Physically, this
corresponds to left and right handed circular waves. We want to allow
for parity transformations exchanging the left and right handed parts,
so we need both parts $\Proj_0 = \Proj_\mathrm{L} + \Proj_\mathrm{R}$.
With the parity transformation included, the two dimensional space
which $\Proj_0$ projects onto is irreducible.

\subsection{Computing the \textit{q}-Maxwell Equations}

We want to write the $q$-Maxwell equations in the form of a first
order differential equation
\begin{equation}
\label{eq:Maxwell2}
  \mathbb{A}\psi := C_\mu P^\mu \, \psi = 0 \,,
\end{equation}
hoping that again the operators $C_\mu$ can be chosen to act on the
spinor index only, $C_\mu = C_\mu \otimes 1$. Recall from the last
section, that as long as we do not include parity transformations, we
must have two independent equations for the right and the left handed
part of the spinor, $\psi_\mathrm{L}$ carrying a $D^{(1,0)}$
representation and $\psi_\mathrm{R}$ carrying a $D^{(0,1)}$
representation
\begin{equation}
  \mathbb{A}_\mathrm{L}\psi_\mathrm{L} = 0 \,,\qquad
  \mathbb{A}_\mathrm{R}\psi_\mathrm{R} = 0 \,.
\end{equation}
For condition~\eqref{eq:WaveCondition1} it would be sufficient (but
not necessary) if $\mathbb{A}_\mathrm{L}$, $\mathbb{A}_\mathrm{R}$
were scalar operators. Let us try to choose $\mathbb{A}_\mathrm{L}=
C_\mu^\mathrm{L} P^\mu$ and $\mathbb{A}_\mathrm{R} = C_\mu^\mathrm{R}
P^\mu$ to be scalars with respect to rotations. For this to be
possible $C_0^\mathrm{L}$, $C_0^\mathrm{R}$ must be scalars with
respect to rotations while $C_A^\mathrm{L}$, $C_A^\mathrm{R}$ must
transform as 3-vectors. The only scalar operators within the
$D^1$ representation of $\suq$ are multiples of the unit matrix, while
every 3-vector operator is proportional to $\rho^1(J_A)$. Hence, up to
an overall constant factor our wave equations would be written as
\begin{equation}
\begin{aligned}
  \bigl(P^0 + \alpha_\mathrm{L}\,\rho^1(J_A)P^A\bigr)\psi_\mathrm{L} &= 0\\
  \bigl(P^0 + \alpha_\mathrm{R}\,\rho^1(J_A)P^A\bigr)\psi_\mathrm{R} &= 0 \,,
\end{aligned}
\end{equation}
where $\alpha_\mathrm{L}$, $\alpha_\mathrm{R}$ are constants. We
determine these constants by considering the wave equations in the
momentum eigenspace,
\begin{xalignat}{2}
  \bigl(1 + \alpha_\mathrm{L}\,\rho^1(J_3)\bigr)\psi_\mathrm{L} &= 0 \,,&
  \bigl(1 + \alpha_\mathrm{R}\,\rho^1(J_3)\bigr)\psi_\mathrm{R} &= 0 \,.
\end{xalignat}
The space of solutions of each of these equations must equal the image
of the projections $\Proj_\mathrm{L}$ and $\Proj_\mathrm{R}$,
respectively. This requirement fixes the constants to
$\alpha_\mathrm{L} = q^{-1}$ and $\alpha_\mathrm{R}= -q$.

Although this determines our candidate for the $q$-Max\-well
equations, condition~\eqref{eq:WaveCondition1} has yet to be checked
for the boosts.  Let $\psi_0$ be an element of the momentum eigenspace
where $(P_0,P_-,P_+,P_3) = (k,0,0,k) =: (p_\mu)$. Using the
commutation relations between boosts and momentum generators we find
\cite{Blohmann}
\begin{subequations}
\begin{xalignat}{2}
  P_\mu (a \psi_0) &= q^{-1} p_\mu(a\psi_0) \,,&
  P_\mu (b \psi_0) &= q^{-1} p_\mu(b\psi_0) \\
  P_\mu (c \psi_0) &= q p_\mu(c\psi_0)  \,,&
  P_\mu (d \psi_0) &= q p_\mu(d\psi_0)\,.
\end{xalignat}
\end{subequations}
By induction it follows, that for any monomial in the boosts, $h=a^i
b^j c^k d^l$, we have $P_\mu (h \psi_0) = q^{k+l-i-j}\,
p_\mu(h\psi_0)$. Thus, for $\psi:=h\psi_0$, the wave
equation~\eqref{eq:Maxwell2} takes the form
\begin{equation}
\label{eq:Maxwell3}
  (C_0 - C_3)\psi = 0 \,.
\end{equation} 
Looking separately at the left and right handed part of $\psi =
\psi_\mathrm{L} + \psi_\mathrm{R}$ this equation writes out
\begin{equation*}
  \begin{pmatrix} 0&0&0 \\0&q^{-2}&0\\0&0&q^{-1}[2]\end{pmatrix}
  \begin{pmatrix} \psi_\mathrm{L}^- \\
                  \psi_\mathrm{L}^3 \\
                  \psi_\mathrm{L}^+ \end{pmatrix} = 0\,,
  \begin{pmatrix} q[2]&0&0 \\0&q^{2}&0\\0&0&0\end{pmatrix}
  \begin{pmatrix} \psi_\mathrm{R}^- \\
                  \psi_\mathrm{R}^3 \\
                  \psi_\mathrm{R}^+ \end{pmatrix} = 0 \,,
\end{equation*}
which is equivalent to $\psi_\mathrm{L}^3 = \psi_\mathrm{L}^+ = 0$ and
$\psi_\mathrm{R}^- = \psi_\mathrm{R}^3 = 0$. If we now have a solution
of Eq.~\eqref{eq:Maxwell3}, that is, a spinor $\psi$ whose only
non-va\-ni\-shing components are $\psi_\mathrm{L}^-$ and
$\psi_\mathrm{R}^+$, could it happen that by boosting it gets other
non-vanishing components, thus turning a solution into a non-solution?
The answer to this question is no. We exemplify this, applying
formula~\eqref{eq:Wave1} for the action of the boost generator $c$ on
a left handed spinor,
\begin{align}
  c\, \psi_\mathrm{L}^A
  &= \rho^{(1,0)}(c_{(1)})^A{}_{A'}
     \bigl(c_{(2)}\tr \psi_\mathrm{L}^{A'}\bigr) \notag\\
  &= \rho^{(1,0)}(c)^A{}_{A'} \bigl(a\tr \psi_\mathrm{L}^{A'}\bigr)
    +\rho^{(1,0)}(d)^A{}_{A'} \bigl(c\tr \psi_\mathrm{L}^{A'}\bigr)\notag\\
  &= -q^{\frac{1}{2}}\lambda \rho^{1}(FK^{\frac{1}{2}})^A{}_{A'}
      \bigl(a\tr \psi_\mathrm{L}^{A'}\bigr)\notag\\
    &\qquad+\rho^{1}(K^{-\frac{1}{2}})^A{}_{A'}
      \bigl(c\tr \psi_\mathrm{L}^{A'}\bigr) \notag\\
  &=\begin{pmatrix}
      -q^{\frac{1}{2}}\lambda[2]^{\frac{1}{2}} a\tr \psi_\mathrm{L}^3
      +q\,c\tr \psi_\mathrm{L}^- \\
      -q^{\frac{1}{2}}\lambda[2]^{\frac{1}{2}} a\tr \psi_\mathrm{L}^+
      +c\tr \psi_\mathrm{L}^3 \\
      q^{-1} c\tr \psi_\mathrm{L}^+
    \end{pmatrix},
\end{align}
which clearly shows that, if $\psi_\mathrm{L}^3$ and
$\psi_\mathrm{L}^+$ vanish, so do $c\psi_\mathrm{L}^3$ and
$c\psi_\mathrm{L}^+$. Similar calculations can be done for the other
boost generators and right handed spinors.

By induction we conclude, that if $\psi_0$ is a solution of
Eq.~\eqref{eq:Maxwell3} and $h=a^i b^j c^k d^l$ is a monomial in the
boosts, the spinor $\psi = h\psi_0$ will be a solution, as well. The
algebra of all boosts, $\SUq^\op$, is generated as linear space by the
monomials, thus, $h\psi_0$ is a solution for any boost $h\in\SUq^\op$.
Since furthermore every $q$-Lorentz transformation can be written as a
sum of products of rotations and boost, $h\psi_0$ is a solution for
\emph{any} $q$-Lorentz transformation $h$. We assume that the space of
solutions, $\ker\mathbb{A}$, is an irreducible representation. This
means in particular that the $q$-Lorentz algebra acts transitively on
$\ker\mathbb{A}$, so any solution can be written as $h\psi_0$.  Hence,
the wave equations
\begin{equation}
\label{eq:Maxwell4}
\begin{aligned}
  \rho^1(J_A)P^A\psi_\mathrm{L} &= -qP_0 \psi_\mathrm{L}\\
  \rho^1(J_A)P^A\psi_\mathrm{R} &= q^{-1}P_0 \psi_\mathrm{R}
\end{aligned} 
\end{equation}
do indeed satisfy property~\eqref{eq:WaveCondition1}. 

Now we want to write these equations as $q$-differential equations
$\tilde{C}_\mu \partial^\mu \psi = 0$, where $\tilde{C}_\mu$ is
defined in Eq.~\eqref{eq:Optilde}.  After lengthy calculations we get
\begin{equation}
\begin{aligned}
  \rho^1(J_{A'})^B{}_{C'}\,\rho^{(1,0)}
  \bigl((L^\Lambda_{\mathrm{I}+})^{A'}{}_{A}\bigr)^{C'}{}_C
  &= -q^2 \varepsilon_C{}^B{}_A  \\
  \rho^1(J_{A'})^B{}_{C'}\,\rho^{(0,1)}
  \bigl((L^\Lambda_{\mathrm{I}+})^{A'}{}_{A}\bigr)^{C'}{}_C
  &= -q^{-2} \varepsilon_C{}^B{}_A \,,
\end{aligned}
\end{equation}
so the wave equations~\eqref{eq:Maxwell4} can be written as
\begin{xalignat}{2}
\label{eq:Maxwell5}
  \vec{\partial}\times \vec{\psi}_\mathrm{L}
  &= \I q^{-1}\partial_0 \vec{\psi}_\mathrm{L}\,, &
  \vec{\partial}\times \vec{\psi}_\mathrm{R}
  &= -\I q\,\partial_0 \vec{\psi}_\mathrm{R}\,,
\end{xalignat}
where $\vec{\psi}_\mathrm{R} = (\psi^A_\mathrm{R})$,
$\vec{\psi}_\mathrm{L} = (\psi^A_\mathrm{L})$ and where the cross
product is defined in Eq.~\eqref{eq:epsidentities3}. A spinor
$\vec{\psi}_\mathrm{L}$ which is a solution to this equation must also
satisfy the mass zero condition. Using
identities~\eqref{eq:epsidentities2} for the cross product,
commutation relations~\eqref{eq:PP-Rel} of the derivations,
$\vec{\partial} \times \vec{\partial} =
-\I\lambda\partial_0\vec{\partial}$, and the wave
equation~\eqref{eq:Maxwell5}, we rewrite the mass zero condition as
\begin{align}
  0 &= \partial_\mu \partial^\mu \vec{\psi}_\mathrm{L}
  = (\partial_0^2 - \vec{\partial}\cdot\vec{\partial})
    \vec{\psi}_\mathrm{L} \notag\\
  &= \partial_0^2 \vec{\psi}_\mathrm{L}
    -(\vec{\partial}\times\vec{\partial})\times \vec{\psi}_\mathrm{L}
    +\vec{\partial}\times(\vec{\partial}\times \vec{\psi}_\mathrm{L})
    -\vec{\partial}(\vec{\partial}\cdot \vec{\psi}_\mathrm{L}) \notag\\
  &=\partial_0^2 \vec{\psi}_\mathrm{L}
    +\I\lambda \partial_0(\vec{\partial} \times \vec{\psi}_\mathrm{L})
    +\vec{\partial}\times(\I q^{-1} \partial_0 \vec{\psi}_\mathrm{L})
    -\vec{\partial}(\vec{\partial}\cdot \vec{\psi}_\mathrm{L}) \notag\\
  &=\partial_0^2 \vec{\psi}_\mathrm{L}
    -q^{-1}\lambda \partial^2_0 \vec{\psi}_\mathrm{L}
    - q^{-2} \partial^2_0 \vec{\psi}_\mathrm{L}
    -\vec{\partial}(\vec{\partial}\cdot \vec{\psi}_\mathrm{L}) \notag\\
  &= -\vec{\partial}(\vec{\partial}\cdot \vec{\psi}_\mathrm{L}) \,. 
\label{eq:Butschi}
\end{align}
Contracting the wave equation with $\vec{\partial}$ 
\begin{equation}
\begin{split}
  \vec{\partial}\cdot(\vec{\partial}\times \vec{\psi}_\mathrm{L})
  &=(\vec{\partial}\times\vec{\partial})\cdot \vec{\psi}_\mathrm{L}
  = -\I\lambda \partial_0 (\vec{\partial}\cdot \vec{\psi}_\mathrm{L}) \\
  &= \I q^{-1}\partial_0 (\vec{\partial}\cdot \vec{\psi}_\mathrm{L})\,,
\end{split}
\end{equation}
we see that $\partial_0 (\vec{\partial}\cdot \vec{\psi}_\mathrm{L}) =
0$ if $\vec{\psi}_\mathrm{L}$ is to satisfy the wave equation.
Together with Eq.~\eqref{eq:Butschi} this means that the mass zero
condition is equivalent to $\partial_\mu (\vec{\partial}\cdot
\vec{\psi}_\mathrm{L}) = 0$, that is, $\vec{\partial}\cdot
\vec{\psi}_\mathrm{L}$ must be a constant number. In a momentum
eigenspace we have $\partial_0 (\vec{\partial}\cdot
\vec{\psi}_\mathrm{L}) = k(\vec{\partial}\cdot
\vec{\psi}_\mathrm{L})$, so this constant number must be zero. The
same reasoning applies to the right handed spinor
$\vec{\psi}_\mathrm{R}$.

We conclude that the wave equations~\eqref{eq:Maxwell5} together with
the mass zero condition $\partial_\mu\partial^\mu \psi = 0$ are
equivalent to 
\begin{xalignat}{2}
\label{eq:Maxwell6}
  \vec{\partial}\times \vec{\psi}_\mathrm{L}
  &= \I q^{-1}\partial_0 \vec{\psi}_\mathrm{L}\,, &
  \vec{\partial}\cdot \vec{\psi}_\mathrm{L} &=0 \\
  \vec{\partial}\times \vec{\psi}_\mathrm{R}
  &= -\I q\,\partial_0 \vec{\psi}_\mathrm{R}\,, &
  \vec{\partial}\cdot \vec{\psi}_\mathrm{R} &=0 \,,
\end{xalignat}
which we will call the $q$-Maxwell equations.

\subsection{The \textit{q}-Electromagnetic Field}

Finally, we write the $q$-Maxwell equations in a more familiar form,
that is, in terms of the $q$-deformed electric and magnetic fields. In
the undeformed case the electric and magnetic fields can --- up to
constant factors --- be characterized within the $D^{(1,0)}\oplus
D^{(0,1)}$ representation as eigenstates of the parity
operator~\eqref{eq:Parity}. The electric field should transform like a
polar vector $\mathcal{P}\vec{E} = - \vec{E}$, while the magnetic
field must be an axial vector $\mathcal{P}\vec{B} = \vec{B}$. Recall,
that the parity operator $\mathcal{P}$ acts on $q$-spinors by
exchanging the left and the right handed parts
$\mathcal{P}\psi_\mathrm{L} = \psi_\mathrm{R}$,
$\mathcal{P}\psi_\mathrm{R} = \psi_\mathrm{L}$.  This fixes the fields
\begin{xalignat}{2}
  \vec{E} &= \I(\vec{\psi}_\mathrm{R} - \vec{\psi}_\mathrm{L}) \,,&
  \vec{B} &= \vec{\psi}_\mathrm{R} + \vec{\psi}_\mathrm{L}
\end{xalignat}
up to constant factors which have been chosen to give the right
undeformed limit. Spinor conjugation of the fields is now the same as
ordinary conjugation $\bar{E}^A = (E^A)^*$, $\bar{B}^A = (B^A)^*$.  In
terms of these fields, the $q$-Maxwell equations~\eqref{eq:Maxwell6}
take the form
\begin{xalignat}{2}
\label{eq:Maxwell7}
  \vec{\partial}\times \vec{E} &= \tfrac{1}{2}[2]\,\partial_0 \vec{B}
     -\tfrac{1}{2}\I\lambda\,\partial_0 \vec{E} \,, &
  \vec{\partial}\cdot \vec{E} &=0 \\
  \vec{\partial}\times \vec{B} &= -\tfrac{1}{2}[2]\,\partial_0 \vec{E}
     -\tfrac{1}{2} \I\lambda \,\partial_0 \vec{B} \,, &   
  \vec{\partial}\cdot \vec{B} &=0 \,.
\end{xalignat}
We can also express the $q$-Maxwell equations in terms of a field
strength tensor $F^{\mu\nu}$. According to the Clebsch-Gordan
series~\eqref{eq:CGseries} we can embed, say, a $D^{(1,0)}$
representation in to a $D^{(\frac{1}{2},\frac{1}{2})}\otimes
D^{(\frac{1}{2},\frac{1}{2})}$ representation. Explicitly, a basis
$\{e_C\}$ of the former is mapped to a basis $\{e'_\mu \otimes e'_\nu\}$
of the latter by \cite{Blohmann}
\begin{equation}
  e_C \mapsto e'_A\otimes e'_B \,\varepsilon^{AB}{}_C
  +q e'_0\otimes e'_C - q^{-1} e'_C\otimes e'_0 \,.
\end{equation}
Accordingly, we map a left $3$-vector
\begin{equation}
  \psi_\mathrm{L} = e_C \otimes  \psi_\mathrm{L}^C \mapsto
  (e'_\mu\otimes e'_\nu)\otimes F_\mathrm{L}^{\mu\nu} \,,
\end{equation}
where
\begin{equation}
  F_\mathrm{L}^{\mu\nu} := \begin{pmatrix}
    F_\mathrm{L}^{00} & F_\mathrm{L}^{0N} \\
    F_\mathrm{L}^{M0} & F_\mathrm{L}^{MN}
  \end{pmatrix}
  = \begin{pmatrix}
    0 & q \psi_\mathrm{L}^N \\
    -q^{-1} \psi_\mathrm{L}^M &
    \varepsilon^{MN}{}_C \, \psi_\mathrm{L}^C
  \end{pmatrix},
\end{equation}
and where $M$, $N$ run through $\{-,+,3\}$. In the same manner we
obtain for the right handed part
\begin{equation}
  F_\mathrm{R}^{\mu\nu} := \begin{pmatrix}
    0 & -q^{-1} \psi_\mathrm{R}^N \\
    q \psi_\mathrm{R}^M & \varepsilon^{MN}{}_C \, \psi_\mathrm{R}^C
  \end{pmatrix}.
\end{equation}
In terms of these matrices the $q$-Maxwell
equations~\eqref{eq:Maxwell6} take the form $\partial_\nu
F_\mathrm{L}^{\mu\nu} = 0$ and $\partial_\nu F_\mathrm{R}^{\mu\nu} =
0$. By construction, we have $\Proj_{\!(1,0)}^{\mu\nu}{}_{\sigma\tau}
F_\mathrm{L}^{\sigma\tau} = F_\mathrm{L}^{\mu\nu}$ and
$\Proj_{\!(0,1)}^{\mu\nu}{}_{\sigma\tau} F_\mathrm{R}^{\sigma\tau} =
F_\mathrm{R}^{\mu\nu}$. This suggests to introduce the field strength
tensor and its dual
\begin{xalignat}{2}
  F^{\mu\nu}
  &:= \I(F_\mathrm{L}^{\mu\nu} + F_\mathrm{R}^{\mu\nu}) \,, &
  \tilde{F}^{\mu\nu}
  &:= \I(F_\mathrm{L}^{\mu\nu} - F_\mathrm{R}^{\mu\nu}) \,, 
\end{xalignat}
for which we have 
\begin{xalignat}{2}
  F^{\mu\nu} &= 
  \Proj_{\mathrm{A}}^{\mu\nu}{}_{\sigma\tau} F^{\sigma\tau}
  \,, & \tilde{F}^{\mu\nu}
  &= \varepsilon^{\mu\nu}{}_{\sigma\tau} F^{\sigma\tau}\,, 
\end{xalignat}
where the $q$-epsilon tensor is commonly defined as
$\varepsilon^{\mu\nu}{}_{\sigma\tau} =
\Proj_{\!(1,0)}^{\mu\nu}{}_{\sigma\tau} -
\Proj_{\!(0,1)}^{\mu\nu}{}_{\sigma\tau}$. In terms of the electric and
the magnetic field this is
\begin{equation}
\begin{aligned}
  F^{\mu\nu} &:= \begin{pmatrix}
    0 & -\tfrac{1}{2}[2]E^N +\tfrac{1}{2}\I\lambda B^N \\
    \tfrac{1}{2}[2]E^M +\tfrac{1}{2}\I\lambda B^M &
    \I\varepsilon^{MN}{}_C \, B^C
  \end{pmatrix} \\
  \tilde{F}^{\mu\nu} &:= \begin{pmatrix}
    0 & \tfrac{1}{2}[2]\I B^N -\tfrac{1}{2}\lambda E^N \\
    -\tfrac{1}{2}[2]\I B^M -\tfrac{1}{2}\lambda E^M &
    -\varepsilon^{MN}{}_C \, E^C
  \end{pmatrix}.
\end{aligned}
\end{equation}
Finally, the $q$-Maxwell equations take the form
\begin{xalignat}{2}
  \partial_\nu F^{\mu\nu} &= 0 \,, &
  \partial_\nu \tilde{F}^{\mu\nu} &= 0 \,,
\end{xalignat}
in complete analogy to the undeformed case.

\section{Conclusion}

In this paper the free Dirac equation and the free Maxwell equations
on quantum Minkowski space have been determined. We started from the
restriction on rest states (or certain light cone states for the
massless case), requiring $q$-Lorentz covariance, effectively
``boosting'' the wave equations. For his method the explicit action of
the momenta on $q$-spinors was not needed and, consequently, was not
constructed. This approach makes the calculations to a large extent
independent of the explicit realization of the momenta. It is in
complete analogy to the classic methods for the undeformed case
\cite{BarutRaczka} and parallels the construction of induced
representations of the Poincar\'e algebra.

The natural next step is to construct the action of the $q$-Poincar\'e
algebra on $q$-spinors, explicitly, as $q$-differential operators on
the $q$-Minkowski algebra of noncommutative wave functions. Then, the
wave equations obtained here can be interpreted as $q$-differential
equations, their solutions can be calculated, and the generalized
Cauchy problem can be studied. This has already been done and will be
published elsewhere.

The free $q$-Dirac equation obtained here is a good starting point for
the construction of the $q$-Dirac equation with electromagnetic
interaction. In fact, the methods to introduce gauge interactions
developed in \cite{Madore:2000b,Jurco:2001} are quite general and do
apply here, although it might be necessary to add a vielbein
formalism \cite{Schraml:2002}. Judging from what is known about models
on spaces of constant noncommutativity (as mentioned in the
introduction) it is not unreasonable to expect the ensuing model on
quantum Minkowski space to possess a rich phenomenology. The
investigation of this phenomenology is expected to be a computational
challenge, though.

On of the main hopes and, at the same time, open questions in
noncommutative geometry is, whether noncommutativity can smoothen the
divergences of quantum field theory. While it would seem bold to claim
that noncommutativity will regularize quantum field theory, there are
some indirect indications that it might weaken the divergences and
improve renormalizability. For example, the spectra of space-time
observables on quantum spaces exhibit a natural lattice structure
\cite{Fichtmuller:1996}. We plan to approach this question by
calculating the Green's functions of the free $q$-wave equations,
since those would appear as bare propagators in loop calculations of
QED on quantum Minkowski space. Studying the divergences of free
propagators on quantum Minkowski space would be a well defined and
rigorous test of the conjectured regularizing properties of
noncommutative geometry.

\subsection*{Acknowledgements}

I would like to thank Julius Wess, Peter Schupp, and Fabian Bachmaier
for helpful comments.  This work was supported by the Studienstiftung
des deut\-schen Volkes.

\appendix

\section{Useful Formulas}
\label{sec:AppPoin}

Let $E$, $F$, $K$, and $K^{-1}$ be the generators of $\suq$.  The set
of generators $\{J_A\} = \{J_-,J_3,J_+\}$ of $\suq$ defined as
\begin{equation}
\label{eq:DefJ}
\begin{aligned}
  J_{-} &:= q[2]^{-\frac{1}{2}}KF \\
  J_3   &:= [2]^{-1} (q^{-1}EF-qFE) \\
  J_{+} &:= -[2]^{-\frac{1}{2}}E
\end{aligned}
\end{equation}
is the left 3-vector operator of angular momentum. The center of
$\suq$ is generated by
\begin{equation}
\label{eq:DefW}
  W := K - \lambda J_3 = K - \lambda [2]^{-1} (q^{-1}EF-qFE)\,,
\end{equation}
the Casimir operator of angular momentum. $W$ is related to $J_A$ by
\begin{equation*}
  W^2 -1 = \lambda^2(J_3^2 - q^{-1} J_- J_+ -  q J_+ J_-)
  = \lambda^2J_A J_B g^{AB} \,,
\end{equation*}
thus defining the 3-metric $g^{AB}$, by which we raise 3-vector
indices $X^A = g^{AB}X_B$. There is also an
$\varepsilon$-tensor
\begin{subequations}
\begin{xalignat}{3}
  \varepsilon^{-3}{}_- &= q^{-1} & \varepsilon^{3-}{}_- &= -q &&\\
  \varepsilon^{-+}{}_3 &= 1 & \varepsilon^{+-}{}_3 &= -1 &
  \varepsilon^{33}{}_3 &= -\lambda \\
  \varepsilon^{3+}{}_+ &= q^{-1} & \varepsilon^{+3}{}_+ &= -q \,,&&
\end{xalignat}
\end{subequations}
so we can define a scalar and a vector product by
\begin{xalignat}{2}
\label{eq:epsidentities3}
  \vec{X}\cdot \vec{Y} &:= g^{AB} X_A Y_B \,,&
  (\vec{X}\times \vec{Y})_C &:= \I\,X_A Y_B\varepsilon^{AB}{}_C \,,
\end{xalignat}
for which we have the useful identities
\begin{equation}
\label{eq:epsidentities2}
\begin{aligned}
  \vec{X} \cdot(\vec{Y} \times \vec{Z})
  &= (\vec{X} \times \vec{Y}) \cdot \vec{Z} \\
  (\vec{X} \times \vec{Y}) \times \vec{Z}
  -(\vec{X}\cdot \vec{Y}) \vec{Z}
  &= \vec{X} \times (\vec{Y} \times \vec{Z})
  -\vec{X} (\vec{Y} \cdot \vec{Z}) \,. 
\end{aligned}
\end{equation}

Let $(\begin{smallmatrix}a&b\\c&d\end{smallmatrix})$ be the matrix of
generators of $\SUq^\op$, the opposite algebra of the quantum group
$\SUq$.  The Hopf-$*$ algebra generated by the Hopf-$*$ subalgebras
$\suq$ and $\SUq^\op$ with cross commutation relations
\begin{equation*}
\begin{aligned}
  \begin{pmatrix} a & b \\ c & d \end{pmatrix} E
  &= \begin{pmatrix} q E a - q^{\frac{3}{2}} b & q^{-1}Eb \\
    qEc+q^{\frac{3}{2}}Ka- q^{\frac{3}{2}}d &
    q^{-1}Ed+q^{-\frac{1}{2}}Kb \end{pmatrix} \\
  \begin{pmatrix} a & b \\ c & d \end{pmatrix} F
  &= \begin{pmatrix} q F a + q^{-\frac{1}{2}}c &
    qFb-q^{-\frac{1}{2}}K^{-1}a + q^{-\frac{1}{2}}d \\
    q^{-1}Fc & q^{-1}Fd-q^{-\frac{5}{2}}K^{-1}c \end{pmatrix} \\
  \begin{pmatrix} a & b \\ c & d \end{pmatrix} K
  &= K \begin{pmatrix} a & q^{-2}b \\ q^{2} c & d\end{pmatrix},
\end{aligned}
\end{equation*}
which is the Drinfeld double of $\suq$ and $\SUq^\op$, is the
\textbf{\textit{q}-Lorentz algebra} $\Hcal = \slC$ \cite{Podles:1990}.

Other forms of the $q$-Lorentz algebra can be found in the literature
\cite{Ogievetskii:1991a,Majid:1993,Lorek:1997a}, which are essentially
equivalent \cite{Rohregger:1999,Blohmann}. Very useful for the
representation theory is the form where $\slC \cong \slq \otimes \slq$
as algebra. This isomorphism is defined on rotations $l\in\suq$ by the
coproduct $l \mapsto \Delta(l) \in \slq \otimes \slq$ and for the
generators of boosts as
\begin{equation}
\label{eq:boostdef}
\begin{aligned}
  a &\mapsto K^{\frac{1}{2}}\otimes K^{-\frac{1}{2}} \\
  b &\mapsto q^{-\frac{1}{2}}\lambda K^{\frac{1}{2}}
    \otimes K^{-\frac{1}{2}} E \\
  c &\mapsto  -q^{\frac{1}{2}}\lambda F K^{\frac{1}{2}}
    \otimes K^{-\frac{1}{2}} \\
  d &\mapsto K^{-\frac{1}{2}}\otimes K^{\frac{1}{2}}
    - \lambda^2 F K^{\frac{1}{2}}\otimes K^{-\frac{1}{2}} E \,.
\end{aligned}
\end{equation}
On this form of the $q$-Lorentz algebra the representation maps of the
irreducible representations are $\rho^{j_1}\otimes\rho^{j_2}$, where
$\rho^j$ is the spin-$j$ representation of $\slq$. For the 4-vector
representation we have the usual Clebsch-Gordan series
\begin{equation}
\label{eq:CGseries}
  D^{(\frac{1}{2},\frac{1}{2})} \otimes D^{(\frac{1}{2},\frac{1}{2})}
  \cong D^{(0,0)} \oplus D^{(1,0)} \oplus D^{(0,1)} \oplus D^{(1,1)} \,.
\end{equation}
The projection matrices on the according subspaces are denoted by
$\Proj_{(0,0)}$, $\Proj_{(1,0)}$, $\Proj_{(0,1)}$, $\Proj_{(1,1)}$,
the $q$-antisymmetrizer by $\Proj_{\mathrm{A}} := \Proj_{(1,0)} +
\Proj_{(0,1)}$, and the $q$-symmetrizer by $\Proj_{\mathrm{S}} :=
\Proj_{(0,0)} + \Proj_{(1,1)} = 1 - \Proj_{\mathrm{A}}$.  With respect
to the 4-vector basis
\begin{center}
\renewcommand{\arraystretch}{1.5}
\begin{tabular}{r|ccc}
\multicolumn{4}{c}{ $[2]^2 (\Proj_{\mathrm{A}})^{ab}{}_{cd}=$ }\\ 
& $C0$ & $0D$  & $CD$ \\ \hline
$A0$ & $2\delta^A_C$ & $-[4][2]^{-1}\delta^A_D$ &
$\lambda \varepsilon_C{}^A{}_D$ \\
$0B$ & $-[4][2]^{-1}\delta^B_C$ & $2\delta^B_D$ &
$\lambda \varepsilon_C{}^B{}_D$ \\
$AB$ & $-\lambda\varepsilon^{AB}{}_C$ &
$-\lambda \varepsilon^{AB}{}_D$ &
$2\varepsilon^{AB}{}_X\varepsilon_C{}^X{}_D$
\end{tabular}
\end{center}
where $A$, $B$, $C$, $D$ run through $\{-,+,3\}$.  $\slC$ possesses
two universal $\R$-matrices, one of which is antireal $\R^{*\otimes *}
= \R^{-1}$, the other one is real $\R^{*\otimes *} = \R_{21}$.

The $*$-algebra generated by $P_0$, $P_-$, $P_+$, $P_3$ with
commutation relations
\begin{xalignat}{2}
\label{eq:PP-Rel}
  P_0 P_A &= P_A P_0 \,,&
  P_A P_B\,\varepsilon^{AB}{}_{C} &= -\lambda P_0 P_C \,, 
\end{xalignat}
and $*$-structure $P_0^* = P_0$, $P_-^* = -q^{-1} P_+$, $P_+^* = -q
P_-$, $P_3^* = P_3$ is the \textbf{\textit{q}-Minkowski space
  algebra} $\Xcal = \Mink$. 
The center of $\Mink$ is generated by
\begin{equation}
\label{eq:FourMetric}
  m^2 := P_\mu P_\nu \eta^{\mu\nu}
  = P_0^2  + q^{-1} P_- P_+ +  q P_+ P_- - P_3^2 \,,
\end{equation}
the mass Casimir, thus defining the 4-metric $\eta^{\mu\nu}$. It is
related to the 3-metric by $\eta^{AB} = - g^{AB}$ for
$A,B\in\{-,+,3\}$. The generators $P_\mu$ carry a 4-vector
representation of $\Hcal$ which in this particular basis is denoted by
$h\tr P_\nu = P_\mu \Lambda(h)^\nu{}_\mu$. It turns $\Xcal$ into a
$\Hcal$-module $*$-algebra.

The \textbf{\textit{q}-Poincar\'e algebra} $\Acal$ is the $*$-algebra
generated by the $q$-Lorentz algebra $\Hcal = \slC$ and the
$q$-Minkowski algebra $\Xcal = \Mink$ with cross commutation relations
\begin{equation}
\label{eq:PoincCommute}
  h\,P_\nu = P_\mu \Lambda(h_{(1)})^\mu{}_\nu \, h_{(2)} \,,
\end{equation}
for all $h\in\Hcal$. In other words, $\Acal$ is the Hopf semidirect
product $\Acal = \Xcal\rtimes\Hcal$.

The left Hopf adjoint action of $\Hcal$ on $\Acal$ is defined as
\begin{equation}
\label{eq:HopfAdjointAction}
  \adL h\tr a := h_{(1)} a S(h_{(2)}) \,.
\end{equation}
The commutation relations~\eqref{eq:PoincCommute} are precisely such
that the left Hopf adjoint action equals the 4-vector action $\adL h
\tr P_\nu = P_\mu \Lambda(h)^\mu{}_\nu$. Any set of operators with
this property will be called a 4-vector operator. In particular we
have
\begin{equation}
\label{eq:Boost1}
\begin{aligned}
 P_- &= \adL(-q^{-\frac{1}{2}}\lambda^{-1}[2]^{\frac{1}{2}} \,c)
             \tr P_0 \\ 
 P_+ &= \adL(q^{\frac{1}{2}}\lambda^{-1}[2]^{\frac{1}{2}} \,b)
             \tr P_0 \\ 
 P_3 &= \adL(\lambda^{-1}\,(d-a)) \tr P_0 \,,
\end{aligned}
\end{equation}
so, if we know the zero component of a 4-vector operator, we can
easily compute the other components.

\section{Gamma matrices for the real R-matrix}

If the real $\R$-matrix instead of the antireal one only the
expressions for the $\gamma$-matrices change slightly.  We now have to
observe that the different projector decomposition of the $R$-matrix
\cite{Lorek:1997a,Blohmann} leads to $\gamma_\mu \gamma_\nu P^\mu
P^\nu = q^3 P_\mu P^\mu$, which differs from the antireal case by the
factor $q^3$. This implies $(\gamma_0)^2 = q^3$. Eq.~\eqref{eq:gamma1}
now becomes
\begin{equation*}
  \gamma_0 =  q^{\frac{3}{2}}
    \begin{pmatrix} 0 & 1 \\ 1 & 0 \end{pmatrix}, \qquad
  \gamma_A = q^{\frac{3}{2}}
    \begin{pmatrix} 0 & q\, \sigma\!_A \\
      -q^{-1}\sigma\!_A & 0 \end{pmatrix} \,,
\end{equation*}
Eq.~\eqref{eq:GammaGeneral} reads
\begin{equation*}
  \gamma_0^{(j)} = q^{\frac{3}{2}}
    \begin{pmatrix} 0 & 1 \\ 1 & 0 \end{pmatrix},\,
  \gamma_A^{(j)} = q^{\frac{3}{2}} [2]
    \begin{pmatrix} 0 & q\,\rho^j(J_A)  \\
    -q^{-1}\rho^j(J_A) & 0 \end{pmatrix},
\end{equation*}
and, finally, Eq.~\eqref{eq:gamma2} has to be replaced by
\begin{equation*}
  \tilde{\gamma}_0 = 
    \begin{pmatrix} 0 & q^{\frac{3}{2}} \\ 
    q^{-\frac{3}{2}} & 0 \end{pmatrix}\,, \qquad
  \tilde{\gamma}_A =
    \begin{pmatrix} 0 & q^{\frac{1}{2}}\, \tilde{\sigma}\!_A \\
      -q^{-\frac{1}{2}} \tilde{\sigma}\!_A & 0 \end{pmatrix} .
\end{equation*}
All other results remain unchanged.

\providecommand{\href}[2]{#2}\begingroup\raggedright\endgroup


\begin{thebibliography}{10}

\bibitem{Douglas:2001}
M.~R. Douglas and N.~A. Nekrasov, ``Noncommutative field theory,'' {\em Rev.
  Mod. Phys.} {\bf 73} (2001) 977--1029,
\href{http://www.arXiv.org/abs/hep-th/0106048}{{\tt hep-th/0106048}}.

\bibitem{Szabo:2001}
R.~J. Szabo, ``Quantum field theory on noncommutative spaces,'' {\em Phys.
  Rept.} {\bf 378} (2003) 207--299,
\href{http://www.arXiv.org/abs/hep-th/0109162}{{\tt hep-th/0109162}}.

\bibitem{Seiberg:1999}
N.~Seiberg and E.~Witten, ``String theory and noncommutative geometry,'' {\em
  JHEP} {\bf 09} (1999) 032,
\href{http://arXiv.org/abs/hep-th/9908142}{{\tt hep-th/9908142}}.

\bibitem{Madore:2000b}
J.~Madore, S.~Schraml, P.~Schupp, and J.~Wess, ``Gauge theory on noncommutative
  spaces,'' {\em Eur. Phys. J.} {\bf C16} (2000) 161--167,
\href{http://www.arXiv.org/abs/hep-th/0001203}{{\tt hep-th/0001203}}.

\bibitem{Jurco:2001}
B.~Jurco, L.~Moller, S.~Schraml, P.~Schupp, and J.~Wess, ``Construction of
  non-Abelian gauge theories on noncommutative spaces,'' {\em Eur. Phys. J.}
  {\bf C21} (2001) 383--388,
\href{http://www.arXiv.org/abs/hep-th/0104153}{{\tt hep-th/0104153}}.

\bibitem{Calmet:2001}
X.~Calmet, B.~Jurco, P.~Schupp, J.~Wess, and M.~Wohlgenannt, ``The standard
  model on non-commutative space-time,'' {\em Eur. Phys. J.} {\bf C23} (2002)
  363--376,
\href{http://www.arXiv.org/abs/hep-ph/0111115}{{\tt hep-ph/0111115}}.

\bibitem{Iltan:2003}
E.~O. Iltan, ``The noncommutative effects on the dipole moments of fermions in
  the standard model,'' {\em JHEP} {\bf 05} (2003) 065,
\href{http://www.arXiv.org/abs/hep-ph/0304097}{{\tt hep-ph/0304097}}.

\bibitem{Minkowski:2003}
P.~Minkowski, P.~Schupp, and J.~Trampetic, ``Non-commutative '*-charge radius'
  and '*-dipole moment' of the neutrino,''
\href{http://www.arXiv.org/abs/hep-th/0302175}{{\tt hep-th/0302175}}.

\bibitem{Schupp:2002}
P.~Schupp, J.~Trampetic, J.~Wess, and G.~Raffelt, ``The photon neutrino
  interaction in non-commutative gauge field theory and astrophysical bounds,''
\href{http://www.arXiv.org/abs/hep-ph/0212292}{{\tt hep-ph/0212292}}.

\bibitem{Abbiendi:2003}
{\bf OPAL} Collaboration, G.~Abbiendi {\em et al.}, ``Test of non-commutative
  QED in the process $\mathrm{e}^+ \mathrm{e}^- \rightarrow \gamma \gamma$ at
  LEP,''
\href{http://www.arXiv.org/abs/hep-ex/0303035}{{\tt hep-ex/0303035}}.

\bibitem{Hinchliffe:2002}
I.~Hinchliffe and N.~Kersting, ``Review of the phenomenology of noncommutative
  geometry,''
\href{http://www.arXiv.org/abs/hep-ph/0205040}{{\tt hep-ph/0205040}}.

\bibitem{Minwalla:1999}
S.~Minwalla, M.~Van~Raamsdonk, and N.~Seiberg, ``Noncommutative perturbative
  dynamics,'' {\em JHEP} {\bf 02} (2000) 020,
\href{http://www.arXiv.org/abs/hep-th/9912072}{{\tt hep-th/9912072}}.

\bibitem{Matusis:2000}
A.~Matusis, L.~Susskind, and N.~Toumbas, ``The IR/UV connection in the
  non-commutative gauge theories,'' {\em JHEP} {\bf 12} (2000) 002,
\href{http://www.arXiv.org/abs/hep-th/0002075}{{\tt hep-th/0002075}}.

\bibitem{Alvarez-Gaume:2003}
L.~Alvarez-Gaume and M.~A. Vazquez-Mozo, ``General properties of noncommutative
  field theories,''
\href{http://www.arXiv.org/abs/hep-th/0305093}{{\tt hep-th/0305093}}.

\bibitem{Bird:1995}
D.~J. Bird {\em et al.}, ``Detection of a cosmic ray with measured energy well
  beyond the expected spectral cutoff due to comic microwave radiation,'' {\em
  Astrophys. J.} {\bf 441} (1995)
144--150.

\bibitem{Amelino-Camelia:2000}
G.~Amelino-Camelia, ``Relativity in space-times with short-distance structure
  governed by an observer-independent (Planckian) length scale,'' {\em Int. J.
  Mod. Phys.} {\bf D11} (2002) 35--60,
\href{http://www.arXiv.org/abs/gr-qc/0012051}{{\tt gr-qc/0012051}}.

\bibitem{Amelino-Camelia:2002}
G.~Amelino-Camelia, ``Doubly-Special Relativity: First Results and Key Open
  Problems,'' {\em Int. J. Mod. Phys.} {\bf D11} (2002) 1643,
\href{http://www.arXiv.org/abs/gr-qc/0210063}{{\tt gr-qc/0210063}}.

\bibitem{Agostini:2003}
A.~Agostini, G.~Amelino-Camelia, and F.~D'Andrea, ``Hopf-algebra description of
  noncommutative-spacetime symmetries,''
\href{http://www.arXiv.org/abs/hep-th/0306013}{{\tt hep-th/0306013}}.

\bibitem{Wachter:2001}
H.~Wachter and M.~Wohlgenannt, ``*-Products on Quantum Spaces,'' {\em Eur.
  Phys. J.} {\bf C23} (2002) 761--767,
\href{http://www.arXiv.org/abs/hep-th/0103120}{{\tt hep-th/0103120}}.

\bibitem{Blohmann:2002a}
C.~Blohmann, ``Covariant realization of quantum spaces as star products by
  Drinfeld twists,'' \href{http://www.arXiv.org/abs/math.qa/0209180}{{\tt
  math.qa/0209180}}.
to appear in J. Math. Phys. \textbf{10} (2003).

\bibitem{Mesref:2002}
L.~Mesref, ``A map between q-deformed noncommutative and ordinary gauge
  theories,''
\href{http://www.arXiv.org/abs/hep-th/0209005}{{\tt hep-th/0209005}}.

\bibitem{Wigner:1939}
E.~P. Wigner, ``On Unitary Representations of the Inhomogeneous Lorentz
  Group,'' {\em Annals Math.} {\bf 40} (1939)
149--204.

\bibitem{Ogievetskii:1992a}
O.~Ogievetskii, W.~B. Schmidke, J.~Wess, and B.~Zumino, ``$q$-Deformed
  Poincar{\'e} algebra,'' {\em Commun. Math. Phys.} {\bf 150} (1992)
495.

\bibitem{Dobrev:1994}
V.~K. Dobrev, ``New q-Minkowski space-time and q-Maxwell equations hierarchy
  from q-conformal invariance,'' {\em Phys. Lett.} {\bf B341} (1994)
133--138.

\bibitem{Schirrmacher:1992}
A.~Schirrmacher, ``Quantum groups, quantum space-time, and Dirac equation,''.
  Talk given at NATO Advanced Research Workshop on Low Dimensional Topology and
  Quantum Field Theory, Cambridge, England, 6-13 Sep 1992.

\bibitem{Pillin:1994b}
M.~Pillin, ``$q$-deformed relativistic wave equations,'' {\em J. Math. Phys.}
  {\bf 35} (1994) 2804--2817,
\href{http://www.arXiv.org/abs/hep-th/9310097}{{\tt hep-th/9310097}}.

\bibitem{Song:1992}
X.-C. Song, ``Covariant differential calculus on quantum Minkowski space and
  the $q$-analog of Dirac equation,'' {\em Z. Phys.} {\bf C55} (1992)
417--422.

\bibitem{Meyer:1995}
U.~Meyer, ``Wave equations on $q$-Minkowski space,'' {\em Commun. Math. Phys.}
  {\bf 174} (1995) 457--476,
\href{http://www.arXiv.org/abs/hep-th/9404054}{{\tt hep-th/9404054}}.

\bibitem{Podles:1996}
P.~Podles, ``Solutions of Klein-Gordon and Dirac equations on quantum Minkowski
  spaces,'' {\em Commun. Math. Phys.} {\bf 181} (1996) 569--586,
\href{http://www.arXiv.org/abs/q-alg/9510019}{{\tt q-alg/9510019}}.

\bibitem{BarutRaczka}
A.~O. Barut and R.~Raczka, {\em Theory of Group Representations and
  Applications}.
\newblock PWN---Polish Scientific Publishers, 1977.

\bibitem{Blohmann}
C.~Blohmann, {\em Spin Representations of the $q$-Poincar{\'e} Algebra}.
\newblock PhD thesis, Ludwig-Maximilians-Universit{\"a}t M{\"unchen}, 2001.
\newblock
\href{http://www.arXiv.org/abs/math.qa/0110219}{{\tt math.qa/0110219}}.
\newblock

\bibitem{Majid:1993}
S.~Majid, ``Braided momentum in the $q$-Poincare group,'' {\em J. Math. Phys.}
  {\bf 34} (1993) 2045--2058,
\href{http://www.arXiv.org/abs/hep-th/9210141}{{\tt hep-th/9210141}}.

\bibitem{Majid}
S.~Majid, {\em Foundations of Quantum Group Theory}.
\newblock Cambridge Univ. Press, 1995.

\bibitem{Blohmann:2001a}
C.~Blohmann, ``Spin in the $q$-Deformed Poincar{\'e} Algebra,''
  \href{http://www.arXiv.org/abs/math.qa/0111008}{{\tt math.qa/0111008}}.
to be published in Commun. Math. Phys.

\bibitem{Schraml:2002}
S.~Schraml, ``Non-Abelian gauge theory on q-quantum spaces,''
\href{http://arXiv.org/abs/hep-th/0208173}{{\tt hep-th/0208173}}.

\bibitem{Fichtmuller:1996}
M.~Fichtm{\"u}ller, A.~Lorek, and J.~Wess, ``$q$-deformed Phase Space and its
  Lattice Structure,'' {\em Z. Phys.} {\bf C71} (1996) 533--538,
\href{http://www.arXiv.org/abs/hep-th/9511106}{{\tt hep-th/9511106}}.

\bibitem{Podles:1990}
P.~Podles and S.~L. Woronowicz, ``Quantum deformation of Lorentz group,'' {\em
  Commun. Math. Phys.} {\bf 130} (1990)
381.

\bibitem{Ogievetskii:1991a}
O.~Ogievetskii, W.~B. Schmidke, J.~Wess, and B.~Zumino, ``Six generator
  $q$-deformed Lorentz algebra,'' {\em Lett. Math. Phys.} {\bf 23} (1991)
233--240.

\bibitem{Lorek:1997a}
A.~Lorek, W.~Weich, and J.~Wess, ``Non-commutative Euclidean and Minkowski
  structures,'' {\em Z. Phys.} {\bf C76} (1997)
375.

\bibitem{Rohregger:1999}
M.~Rohregger and J.~Wess, ``$q$-deformed Lorentz-algebra in Minkowski phase
  space,'' {\em Eur. Phys. J.} {\bf C7} (1999), no.~1, 177--183.

\end{thebibliography}
\end{document}